\documentclass[aps,pra,superscriptaddress,showkeys,showpacs,nofootinbib,twocolumn]{revtex4-1}
\usepackage{amsmath}    
\usepackage{graphicx}   
\usepackage{array}
\usepackage[dvipdfx,cmyk,natbib]{xcolor}     
\usepackage[caption=false]{subfig}  
\usepackage[colorlinks,linkcolor=red,citecolor=brown,urlcolor=blue]{hyperref}
\usepackage{dsfont,bm}
\usepackage{amssymb}

\newcommand{\be}{\begin{equation}}
\newcommand{\ee}{\end{equation}}
\newcommand{\ben}{\begin{eqnarray}}
\newcommand{\een}{\end{eqnarray}}
\newcommand{\bes}{\begin{subequations}}
\newcommand{\ees}{\end{subequations}}
\newcommand{\bF}{\begin{figure}}
\newcommand{\eF}{\end{figure}}

\newcommand{\ttt}[1]{\text{#1}}
\newcommand{\bra}[1]{\langle{#1}\vert}
\newcommand{\ket}[1]{\vert{#1}\rangle}

\newcommand{\avg}[1]{\langle {#1} \rangle}

\begin{document}

\title{Boson sampling with Gaussian measurements}

\author{L. Chakhmakhchyan}
\author{N. J. Cerf}
\affiliation{Centre for Quantum Information and Communication, Ecole polytechnique de Bruxelles,
CP 165, Universit\'{e} libre de Bruxelles, 1050 Brussels, Belgium}

\date{\today}
\begin{abstract}
	
We develop an alternative boson sampling model operating on single-photon states followed by linear interferometry and Gaussian measurements. The hardness proof for simulating such continuous-variable measurements is established in two main steps, making use of the symmetry of quantum evolution under time reversal. Namely, we first construct a twofold version of scattershot boson sampling in which, as opposed to the original proposal, both legs of a collection of two-mode squeezed vacuum states undergo parallel linear-optical transformations. This twofold scattershot model yields, as a corollary, an instance of boson sampling from Gaussian states where photon counting is hard to simulate. Then, a time-reversed setup is used to exhibit a boson sampling model in which the simulation of Gaussian measurements -- namely the outcome of eight-port homodyne detection -- is proven to be computationally hard. These results illustrate how the symmetry of quantum evolution under time reversal may serve as a tool for analyzing the computational complexity of novel physically-motivated computational problems.

\end{abstract}

\maketitle

\section{Introduction}

Ever since the field of quantum computation emerged, the construction of a large-scale  universal quantum computer has been an ultimate experimental challenge. However, despite remarkable technological progress, quantum computational primitives are still demonstrated at a proof-of-principle level in all proposed implementations today~\cite{tasks}. In an attempt to overcome this state of affairs, several schemes have been developed for technically less demanding but non-universal quantum devices, which are nevertheless usable for demonstrating the power of quantum computers~\cite{restr}. Specifically, one such model of a restricted quantum computer which realizes the task of boson sampling was proposed by Aaronson and Arkhipov~\cite{boson}. In its photonic implementation, identical single photons are injected into a multimode linear-optical interferometer and the task is to sample from the output photon-detection probability distribution. Given plausible complexity-theoretic assumptions, the hardness of this task originates from the $\#$P computational complexity of matrix permanents~\cite{boson}, which enter the expression of photon-counting probabilities.

Currently, boson sampling is viewed as one of the platforms which may enable to  demonstrate the advantage of quantum computers, a promise which has motivated various small-scale realizations~\cite{bs}. Nevertheless, the scaling of boson sampling experiments still faces challenges, so that the near-term demonstration of the quantum advantage via boson sampling remains unlikely~\cite{imm, imm1}. For example, the efficient generation of single-photon input states represents a primary limitation, mainly due to the probabilistic nature of the most common single-photon sources. As a result, several variants of boson sampling have been proposed in order to facilitate its implementation~\cite{ralph, spin}, as specifically the scattershot boson sampling~\cite{scattershot}. The latter model relies on two-mode squeezed vacuum states (TMSSs) where one mode of each state is used for heralding the presence (at random positions) of single photons, while the other set of modes undergo a linear-optical transformation. This randomization of the input preparation step enables to achieve an exponential improvement in the single-photon generation rate, as compared to the original (fixed-input) boson sampling scheme. Moreover, the setup can be seen as a specific instance of boson sampling from Gaussian states: one samples the output photon-counting probability starting from a collection of TMSSs at the input, with one half of the modes going through a linear interferometer and the other half remaining unchanged (heralding can be seen as participating in the output photon counting).

In this paper, we start by extending the scattershot boson sampling paradigm. Namely, we consider a model where both legs of a collection of TMSSs undergo a linear-optical unitary transformation, respectively U$_\ttt{A}$ and U$_\ttt{B}$ [see Fig.~\ref{scatt}(a)]: the original scattershot setup can be recovered from this scheme simply by replacing one of the unitaries with the identity. Inspired by the retrodictive approach of quantum mechanics~\cite{retro}, we show that such an extended model, which we call {\it twofold} scattershot boson sampling (TSBS), can be reduced to the original boson sampling by Aaranson and Arkhipov provided that all TMSSs are equally squeezed. The hardness of the TSBS  therefore follows.

Furthermore, the TSBS  can be seen as boson sampling starting with a collection of pairs of single-mode squeezed vacuum states, which are combined pairwise on beam splitters and processed in two parallel linear-optical unitary transformations (see Fig.~\ref{squeezed}). Our proof thus enlarges the set of known classically-hard sampling tasks with squeezed vacuum states, which was up to now restricted to the corresponding analogue of scattershot boson sampling (where one of the unitaries is the identity). Interestingly, a more general result was recently reported in Ref.~\cite{GBS}, where the model of Gaussian boson sampling was introduced (the current work was carried out independently). This model involves, specifically, a collection of squeezed states at the input, while the corresponding photon-counting probabilities are defined in terms of hafnians~\cite{hafnian} of complex matrices. The hardness proof reported in Ref.~\cite{GBS} shares the equal squeezing requirement with our approach (its role becomes more evident in our approach), but, unlike our proof, it relies on the computational complexity of hafnians.

Building on this TSBS scheme, we go on with time symmetry considerations and define a novel model of boson sampling involving single-photon input states and Gaussian measurements at the output of a linear interferometer. It can be understood as the time-reversed version of TSBS, where Gaussian input states are processed via a linear interferometer and sampled by means of single-photon detectors. The Gaussian and non-Gaussian components are simply interchanged here. Our complexity analysis is based on the aforementioned hardness proof of sampling from squeezed vacuum states in our TSBS model, thus illustrating how time symmetry motivates the analysis of new computational problems and is a useful tool for assessing their complexity. More specifically, we show that projecting linear-optically evolved single-photon states on displaced squeezed states (which is a Gaussian measurement achieved by eight-port homodyne detection~\cite{homo1,ulf}; see 
Fig.~\ref{gaussian}) yields a continuous-variable probability density that is hard to sample. Notably, this construction addresses an open problem discussed in Ref.~\cite{boson} as it constitutes an explicit instance of boson sampling with Gaussian measurements. It further explores the minimal extensions to Gaussian quantum computational models~\cite{knill} that are needed to enable a quantum advantage: previous approaches either required non-Gaussian evolution, or made use of Gaussian input states and evolution but non-Gaussian measurements~\cite{ralph, GBS, scattershot, CVI}. In contrast, our model operates on non-Gaussian input states, but needs only Gaussian evolution (esp. a linear interferometer) and Gaussian measurements.

Finally, let us stress that in this paper we deal with the task of {\it exact} boson sampling, which consists of sampling from the exact output probability distribution. However, real experiments have imperfections that make this distribution to deviate slightly from the ideal model, leading to the issue of noise. This motivated the definition of the {\it approximate} boson sampling model which consists of sampling from a probability distribution that is constrained in variation distance to the exact one~\cite{boson}. Since then, various forms of noise in a boson sampling experiment (with single-photon inputs and photon counting) that preserve its classical computational hardness have been analyzed. For example, sufficient level of tolerance for beam splitters and phase shifters composing the linear optical network~\cite{alex}, lower limit on the required indistinguishability of photons to achieve the quantum advantage~\cite{oxf} and effects of photon losses were reported~\cite{photon}. Further, the amount of Gaussian error applied to the overall unitary matrix~\cite{kilai} and sufficient conditions for classically simulating the boson sampling experiment (in terms of quasiprobability distribution functions) were also addressed~\cite{prx}. The case of approximate boson sampling with Gaussian measurements will be analyzed in a further work.

The paper is organized as follows. In Section~\ref{GSBS}, we describe the model of twofold scattershot boson sampling (TSBS) and prove its hardness. In Section~\ref{new}, we construct a boson sampling model with Gaussian measurements, thereby providing the first example of a computationally hard task consisting of sampling from a continuous-variable probability density. Finally, we draw our conclusions in Sec.~\ref{concl}.

\section{Twofold  scattershot boson sampling}\label{GSBS}
\subsection{Reduction to original boson sampling}\label{TMSS}

\begin{figure}[h!]
\begin{center}
(a) \includegraphics[width=0.45\textwidth]{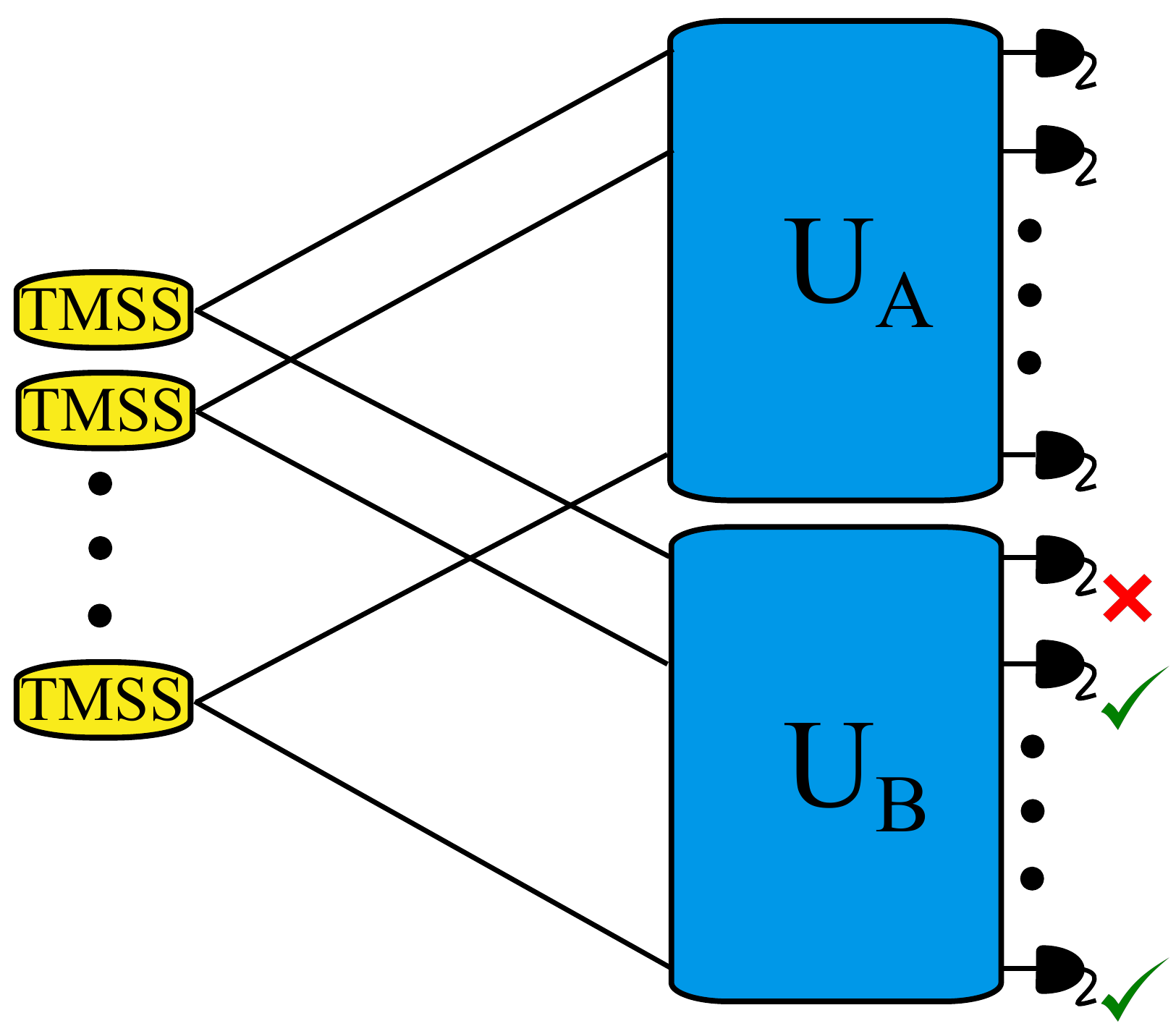}\\
(b) \includegraphics[width=0.45\textwidth]{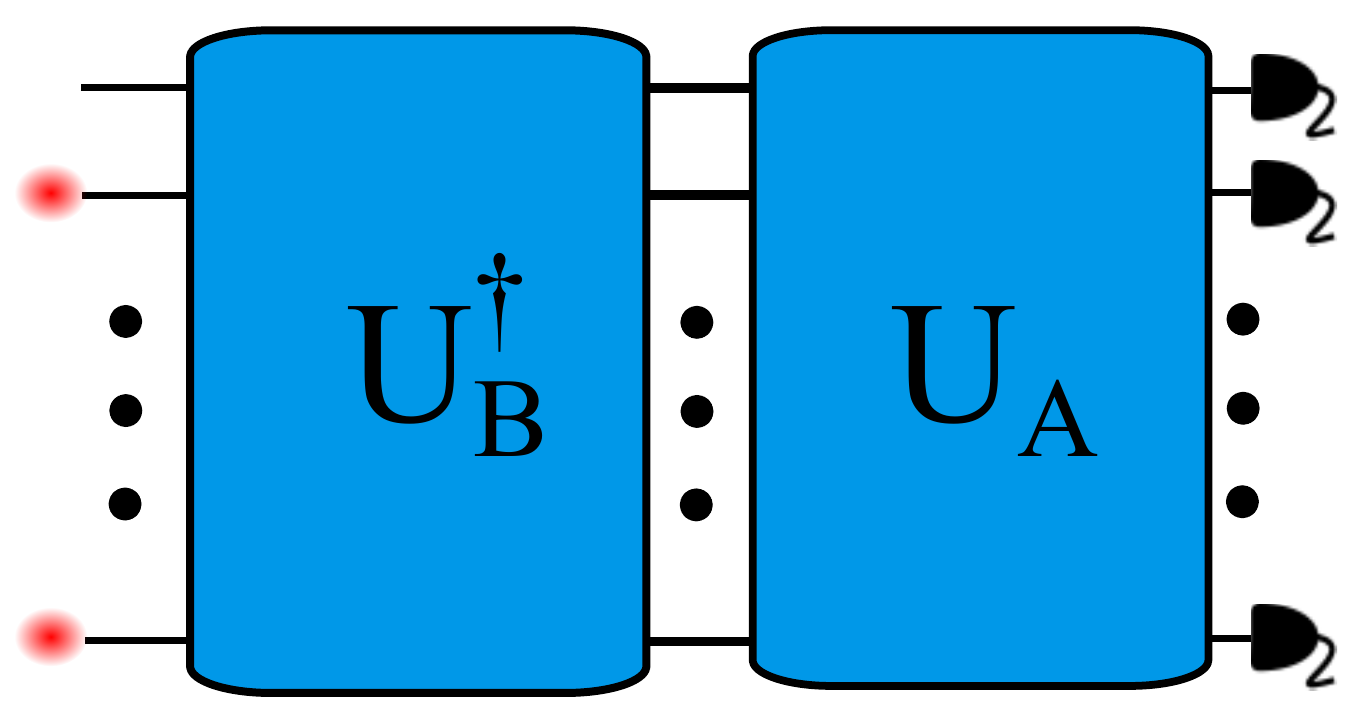}
\caption {(a) Twofold scattershot boson sampling setting. The input is a set of $M$ TMSSs. One half of the $i$th TMSS is sent to the $i$th mode of the linear-optical circuit U$_\ttt{A}$, while the second half is injected into the $i$th mode of the optical circuit U$_\ttt{B}$. A sample of single photon detections is acquired at the output of U$_\ttt{A}$, upon the postselection of detecting a specific pattern of single photons at output port of U$_\ttt{B}$; (b) The equivalent partially time-reversed (time-unfolded) setup of (a), interpreted as boson sampling with single-photon input states. \label{scatt}}
\end{center}
\end{figure}

We start by modifying the scattershot boson sampling model proposed in Ref.~\cite{scattershot}. We consider a set of $M$ two-mode  squeezed vacuum states (TMSSs) $\ket{\psi_\ttt{in}}=\otimes_{j=1}^M\ket{\psi_j}$. Each TMSS is characterized in terms of its squeezing parameter $0\leq t_j<1$ and can be written down as $\ket{\psi_j}=\sqrt{1-t_j^2}\sum_{n_i=0}^\infty t_j^{n_i}\ket{n_i}_\ttt{A}\ket{n_i}_\ttt{B}$, where A and B label the two sides. One leg of the $j$th TMSS is then injected into the $j$th mode (out of $M$)  of the linear-optical circuit U$_\ttt{A}$ and the second leg is sent to the $j$th mode of U$_\ttt{B}$ [see also Fig.~\ref{scatt}(a)]. We use here the description of the circuit in terms of two $M\times M$ unitary matrices, which transform the input mode operators $\hat{a}_l^{(i)^\dagger}$ (with $l=1,\dots, M$) onto the output mode operators $\hat{b}_k^{(i)^\dagger}$ (with $k=1,\dots, M$):
\begin{equation}\label{1}
\hat{b}_k^{(i)^\dagger}=\sum_{j=1}^M {\ttt U}^{(i)}_{{kl}} \, \hat{a}_l^{(i)^\dagger}, \,\,\,\, i=\ttt{A}, \ttt{B}.
\end{equation}
Remark that there is a natural homomorphism between the $M\times M$ unitary matrix $\ttt{U}_i$ (whose matrix elements are noted ${\ttt U}^{(i)}_{{kl}}$) and the corresponding unitary transformation $\mathcal{U}_i$ in state space with $\ket{\psi_\ttt{out}}=\mathcal{U}_\ttt{A}\otimes \mathcal{U}_\ttt{B}\ket{\psi_\ttt{in}}$, so we will use these two descriptions of a linear-optical circuit  interchangeably. 

Assume that a specific pattern ${\bf m}\equiv \{m_1, \dots, m_M\}$ of $N$ single photons ($\sum_{i=1}^M m_i=N\leq M$, with $m_i\in \{0,1\}$, $\forall i$) has been detected at the output of the circuit U$_\ttt{B}$. As we shall show, the problem of sampling the single-photon detection probability distribution at the output of the circuit U$_\ttt{A}$ conditionally on ${\bf m}$ is a computationally hard task, which we call twofold scattershot boson sampling (TSBS). The computational hardness of TSBS follows from the fact that it can be reduced to an instance of the original boson sampling model defined in Ref.~\cite{boson}. 
In order to prove this statement, we write down the conditional probability of detecting a pattern ${\bf k}\equiv \{k_1, \dots, k_M\}$ of single photons  at the output of U$_\ttt{A}$, given the detection of pattern ${\bf m}$ at the output of U$_\ttt{B}$:

\ben\label{2.1} \nonumber
p({\bf k}| {\bf m})=\frac{p({\bf k}\cap {\bf m})}{p({\bf m})}=\frac{p({\bf k}\cap {\bf m})}{\sum_{{k_1, \dots, k_M}}p({\bf k}\cap {\bf m})}.
\een
Here $p({\bf k}\cap {\bf m})$ is the joint probability of detecting the state $\ket{{\bf k}}_\ttt{A}$ at the output of U$_\ttt{A}$ and the state $\ket{{\bf m}}_\ttt{B}$ at the output of U$_\ttt{B}$,

\ben\label{2} \nonumber
p({\bf k}\cap {\bf m})= \left|{}_\ttt{A}\bra{{\bf k}} _\ttt{B}\avg{{\bf m}|\psi_\ttt{out}}\right|^2=\prod_{i=1}^M(1-t_i^2)\\ 
\times \left|\sum_{{n_1, \dots, n_M}=0}^{\infty}t_1^{n_1}\cdots t_M^{n_M} {}_\ttt{A}\avg{{\bf k}|\mathcal{U}_\ttt{A}|{\bf n}}_\ttt{A}\cdot
{}_\ttt{B}\avg{{\bf m}|\mathcal{U}_\ttt{B}|{\bf n}}_\ttt{B}\right|^2,
\een
where ${\bf n}\equiv \{n_1, \dots, n_M\}$. Note that since we operate with linear optics, given the detection of $N$ single photons at the output of U$_\ttt{B}$, the same number of photons must emerge from U$_\ttt{A}$. Thus we have the constraint $\sum_{i=1}^Mk_i=\sum_{i=1}^M n_i =N$.  Next, we assume that the squeezing parameters of TMSSs $\ket{\psi_i}$ are all equal, i.e., $t_1=\dots=t_M\equiv t$, which, as explained below, is vital to our proof. Consequently, Eq.~(\ref{2}) is rewritten as
\ben\label{3} \nonumber
p({\bf k}\cap {\bf m})&&=(1-t^2)^{M}t^{2N} \\
&&\times \left|\sum_{\sum_{i=1}^M n_i=N} \avg{{\bf k}|\mathcal{U}_\ttt{A}|{\bf n}}
\avg{{\bf m}|\mathcal{U}_\ttt{B}|{\bf n}}\right|^2,
\een
where the summation runs over all patterns of $n_i$'s summing up to $N$, and for simplicity, the subscripts A and B of the Fock states have been omitted. Next, 
\ben\label{3.1} \nonumber
p({\bf m})=\sum_{k_1, \dots, k_M}\left|{}_\ttt{A}\bra{{\bf k}} _\ttt{B}\avg{{\bf m}|\psi_\ttt{out}}\right|^2=\avg{{\bf m}|\mathcal{U}_\ttt{B}(\ttt{Tr}_\ttt{A} \rho_{\ttt{in}})\mathcal{U}_\ttt{B}^\dagger|{\bf m}},
\een
where $\ttt{Tr}_\ttt{A} \rho_{\ttt{in}}\equiv \ttt{Tr}_\ttt{A} \ket{\psi_{\ttt{in}}}\bra{\psi_{\ttt{in}}}$ denotes the state obtained after tracing out $\ket{\psi_{\ttt{in}}}$ over subsystem A. Since $\ket{\psi_{\ttt{in}}}$ is a collection of $M$ TMSSs, $\ttt{Tr}_\ttt{A} \rho_{\ttt{in}}$ is a $M$-mode thermal state, 
\be \label{3.2}
\ttt{Tr}_\ttt{A} \rho_{\ttt{in}}=\bigotimes_{j=1}^M\left[(1-t_j^2)\sum_{n_i=0}^{\infty}t_j^{2 n_i}\ket{n_i}\bra{n_i}\right].
\ee
Taking the equal squeezing assumption into account and remembering that a linear-optical transformation leaves a $M$-mode thermal states of equal mean photon numbers ($t_j\equiv t, \forall j$) unchanged, we find $p({\bf m})=(1-t^2)^{M}t^{2N}$, yielding
\be \label{3.3}
p({\bf k}|{\bf m})=\left|\sum_{\sum_{i=1}^M n_i=N} \avg{{\bf k}|\mathcal{U}_\ttt{A}|{\bf n}}
\avg{{\bf m}|\mathcal{U}_\ttt{B}|{\bf n}}\right|^2.
\ee

We now show that the task of sampling from the probability distribution $p({\bf k}| {\bf m})$, which we denote as TSBS, reduces to the original boson sampling problem. The latter emerges here as a partially time-reversed version of the TSBS. Namely, we unfold the TSBS setup by back-propagating in time the state of the  modes on $B$ side. This partially time-reversed or time-unfolded TSBS results in a boson sampling setting with a multimode Fock state $\ket{{\bf m}}$ at the input which is processed into two consecutive linear-optical circuits, U$_\ttt{B}^{\dagger}$ and U$_\ttt{A}$, as depicted in Fig.~\ref{scatt}(b)  (the Hermitian conjugation of the matrix U$_\ttt{B}$ represents time reversal, corresponding in turn to the transposition of $\mathcal{U}_\ttt{B}$ in the Fock basis).
Now, the conditional probability of detecting a pattern {\bf k} of single photons at the output of the described scattering process, given the input state $\ket{{\bf m}}$, reads:
\ben\nonumber \label{5}
\tilde{p}({\bf k}| {\bf m})=&&|\avg{{\bf k}|\mathcal{U}_\ttt{A}\mathcal{U}_\ttt{B}^\ttt{T}|{\bf m}}|^2=\left|\sum_{\sum_{i=1}^Mn_i=N}\bra{{\bf k}}\mathcal{U}_\ttt{A} \ket{{\bf n}}\avg{{\bf n}|\mathcal{U}_\ttt{B}^\ttt{T}|{\bf m}}\right|^2 \\ 
&& =\left|\sum_{\sum_{i=1}^Mn_i=N}\bra{{\bf k}}\mathcal{U}_\ttt{A} \ket{{\bf n}}\avg{{\bf m}|\mathcal{U}_\ttt{B}|{\bf n}}\right|^2.
\een
which coincides with ${p}({\bf k}| {\bf m})$. This confirms that the twofold boson sampling setup [Fig.~\ref{scatt}(a)] is formally equivalent to the original boson sampling setup [Fig.~\ref{scatt}(b)].
Hence, the computational hardness proof for the exact sampling task~\cite{boson} is directly applicable, since the two consecutive linear-optical unitary transformations U$_\ttt{B}^\dagger$ and U$_\ttt{A}$ can be combined into a single one, defined in terms of the unitary matrix U=U$_\ttt{A}$U$_\ttt{B}^\dagger$ and
\ben\label{5.1}
\tilde{p}({\bf k}| {\bf m})=\frac{|\ttt{Perm U}_{{\bf k}, {\bf m}}|^2}{k_1!\cdots k_M!},
\een
where $\ttt{Perm U}_{{\bf k}, {\bf m}}$ stands for the permanent of the $N\times N$ matrix U$_{{\bf k}, {\bf m}}$, which is obtained from the unitary U by deleting its $i$th column if $m_i=0$ and repeating $k_j$ times its $j$th row (or deleting the $j$th row if $k_j=0$).

Furthermore, the computational complexity of the approximate sampling task can also be proven by following Ref.~\cite{boson} and adding the requirement that the two unitary transformations U$_\ttt{B}^\dagger$ and U$_\ttt{A}$ are drawn from the Haar measure, so that their product, U, is again a Haar random unitary. One has to satisfy, however, yet another condition on the ratio between the number of single photons $N$ and the size of the circuit $M$. Strictly speaking, the hardness proof for the original approximate boson sampling holds if $M>N^6$. However, it is conjectured in Ref.~\cite{boson} that the complexity arguments would hold even when $M=O(N^2)$. Consequently, to ensure that the hardness proof of the TSBS model holds, one has to perform postselection upon the events where the number $N$ of photons detected at the output of U$_\ttt{B}$ satisfy $M=O(N^2)$.

Note that the original scattershot boson sampling setting proposed in Ref.~\cite{scattershot} corresponds obviously to a special case of TSBS with U$_\ttt{B}$ being the $M\times M$ identity matrix. Importantly, in our setting, postselection can be performed upon the outputs of either circuit U$_\ttt{B}$ or U$_\ttt{A}$, while in the original scheme~\cite{scattershot} it must only take place upon the outputs of the identity channel.

Remark that the equal squeezing of the collection of input TMSSs is crucial to our proof, specifically to the derivation of Eqs.~(\ref{3}) and~(\ref{3.3}). Namely, this condition allows one to factor out the squeezing dependence from the joint probability $p({\bf k}\cap {\bf m})$, which then appears as a constant term in Eq.~(\ref{3}), independent of the measurement outcome. This is a key point in the boson sampling model from Gaussian states described in Sec.~\ref{squeezedd}.
Provided this condition holds, the joint probability distribution $p({\bf k}\cap {\bf m})$ in the TSBS setup differs from $\tilde{p}({\bf k}|{\bf m})$ in the original boson sampling model by a constant prefactor, $p({\bf m})=(1-t^2)^M t^{2N}$, which depends on the squeezing parameter of the input states (notably, the same holds for scattershot boson sampling, too~\cite{scattershot}). The presence of this prefactor, however, does not affect the hardness proof, in analogy with the situation for the scattershot boson sampling. 
In other words, given the hardness of sampling $\tilde{p}({\bf k}|{\bf m})$, sampling the joint distribution $p({\bf k}\cap {\bf m})$ represents a computationally hard problem as well. Moreover, 
following Ref.~\cite{scattershot}, we conclude that when employing two-mode squeezed vacuum states instead of single-photon input states, the TSBS has the same advantage as the scattershot boson sampling with respect to the original boson sampling. And specifically, the optimal squeezing degree $t_\ttt{opt}$ that maximizes this gain is achieved at $t_\ttt{opt}=1/\sqrt{M+1}$ (for $N=M^2$).

Finally, we note that the above equivalence between the TSBS and the original boson sampling model can be understood from a simple perspective. Namely, given the equal squeezing condition of the $M$ TMSSs, a detection of a pattern {\bf m} of $N$ single photons at one of its legs yields a discrete-variable ${M}\choose{N}$-dimensional maximally entangled Bell state. In turn, a linear-optical unitary transformation at one of its sides is equivalent to applying the transposed unitary on the other side. Consequently, the TSBS is reduced to the original scattershot boson sampling scheme, where pattern {\bf m} is used to herald the input of a standard (fixed-input) boson sampling model.

\subsection{Boson sampling with squeezed vacuum states}\label{squeezedd}

\begin{figure}[h!]
	\begin{center}
		\includegraphics[width=0.45\textwidth]{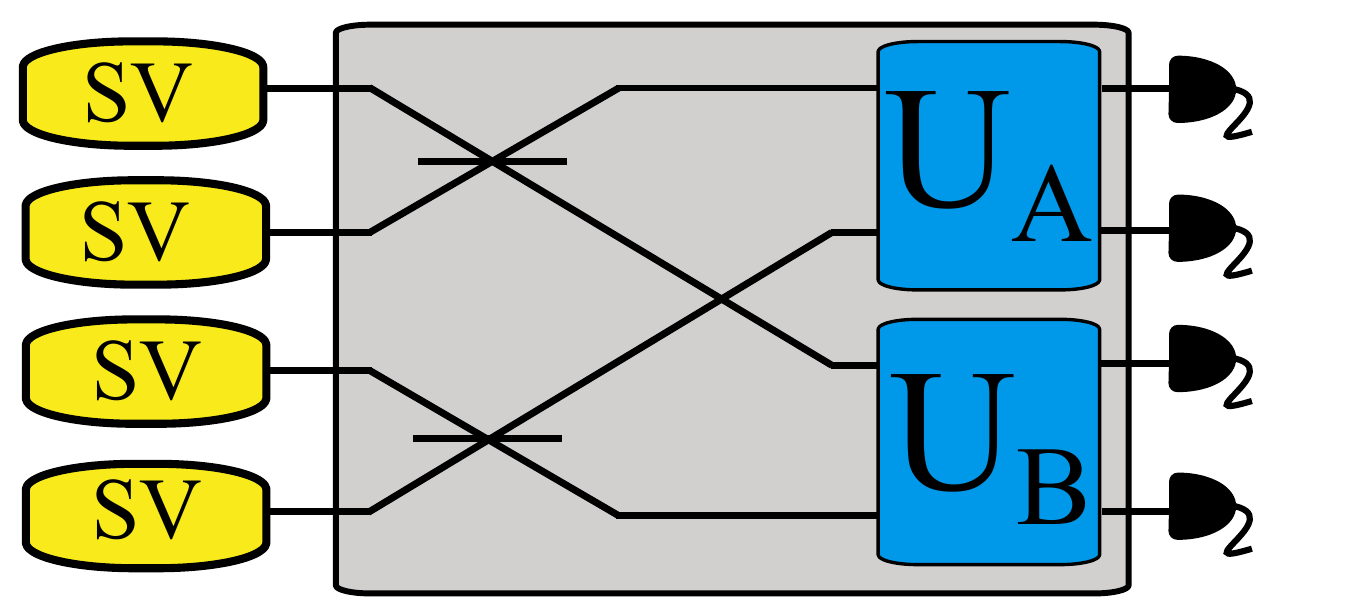}
		\caption {A four-mode example of computationally hard boson sampling setting with squeezed vacuum (SV) input states, equivalent to twofold scattershot boson sampling. The shaded square stands for the linear optical transformation acing upon the SV states.  \label{squeezed}}
	\end{center}
\end{figure}

The twofold scattershot boson sampling model yields, as a straightforward corollary, a result on the computational hardness of boson sampling with squeezed vacuum input states. First, we note that a TMSS of squeezing $t$ can be generated by mixing two squeezed vacuum states $\ket{\xi}$ and $\ket{-\xi}$ on a balanced beam splitter ($\xi$ is the squeezing parameter of the state, $t=\tanh \xi$). Therefore, the setup depicted in Fig.~\ref{scatt}(a) can be considered as a special instance of boson sampling with $2M$ squeezed vacuum input states, which are then mixed by pairs on $M$ beam splitters. Next, the collection of ``upper" and ``lower"  modes  are sent to two linear optical circuits U$_\ttt{A}$ and U$_\ttt{B}$, resulting in the overall $2M$-mode unitary (see Fig.~\ref{squeezed} for a four-mode example)  
\be \label{5.1.1}
\ttt{U}\equiv(\ttt{U}_\ttt{A}\oplus \ttt{U}_\ttt{B}) \cdot (\oplus_{i=1}^{M}\ttt{U}_\ttt{BS}),
\ee
where 
\be\label{6}
\ttt{U}_\ttt{BS}=
\frac{1}{\sqrt{2}}\begin{bmatrix}
	1 & 1  \\
	-1 & 1 
\end{bmatrix}
\ee
is the unitary effected by the balanced beam splitter transformation. As shown in the previous subsection, sampling (exactly or approximately) from the probability distribution of detecting a pattern {\bf k} of $N$ photons at the output of circuit U$_\ttt{A}$, conditioned on the detection of pattern {\bf m} at the output of U$_\ttt{B}$, represents a computationally hard problem (given the conditions on $N$ and $M$). Consequently, provided the equal-squeezing condition is fulfilled, sampling from the joint probability distribution of detecting pattern {\bf k} at the output of U$_\ttt{A}$ and pattern {\bf m} at the output of U$_\ttt{B}$ is computationally hard too. More precisely, the probability distribution from which one samples here is, according to Eqs.~(\ref{3.3}), (\ref{5}) and (\ref{5.1}),
\be\label{5.2}
p_{\pmb{\xi}_0}({\bf k} \cap {\bf m}) = (1-t^2)^M t^{2N}  \frac{|\ttt{Perm U}_{{\bf k}, {\bf m}}|^2}{k_1!\cdots k_M!},
\ee
where the subscript $\pmb{\xi}_0\equiv \{\xi, -\xi, \dots, \xi, -\xi\}$  denotes the vector of squeezing parameters of the $2M$ input states.

Note that a more general result for boson sampling with squeezed vacuum input states was independently reported in the recent Ref.~\cite{GBS}, where the model of Gaussian boson sampling was introduced. As opposed to our approach, the  proof of Ref.~\cite{GBS} relies on the computational complexity of hafnians, as emerging in the corresponding photon-counting probabilities. It thus implies our result as a special case (incidentally, we note that the proof of Ref.~\cite{GBS} requires the equal squeezing of the input states as well). While they are based on distinct approaches, Ref.~\cite{GBS} and our result extend the paradigm of boson sampling to the task of simulating photon counting at the output of a linear interferometer with Gaussian input states (squeezed states).

\section{Time-reversed model with Gaussian measurements}\label{new}

Exploiting the symmetry of quantum evolution under time reversal, we now proceed with constructing a model of boson sampling with Gaussian measurements (i.e., measurements whose POVM elements are projectors onto Gaussian pure states). 
\begin{figure}[h!]
	\begin{center}
		\includegraphics[width=0.4\textwidth]{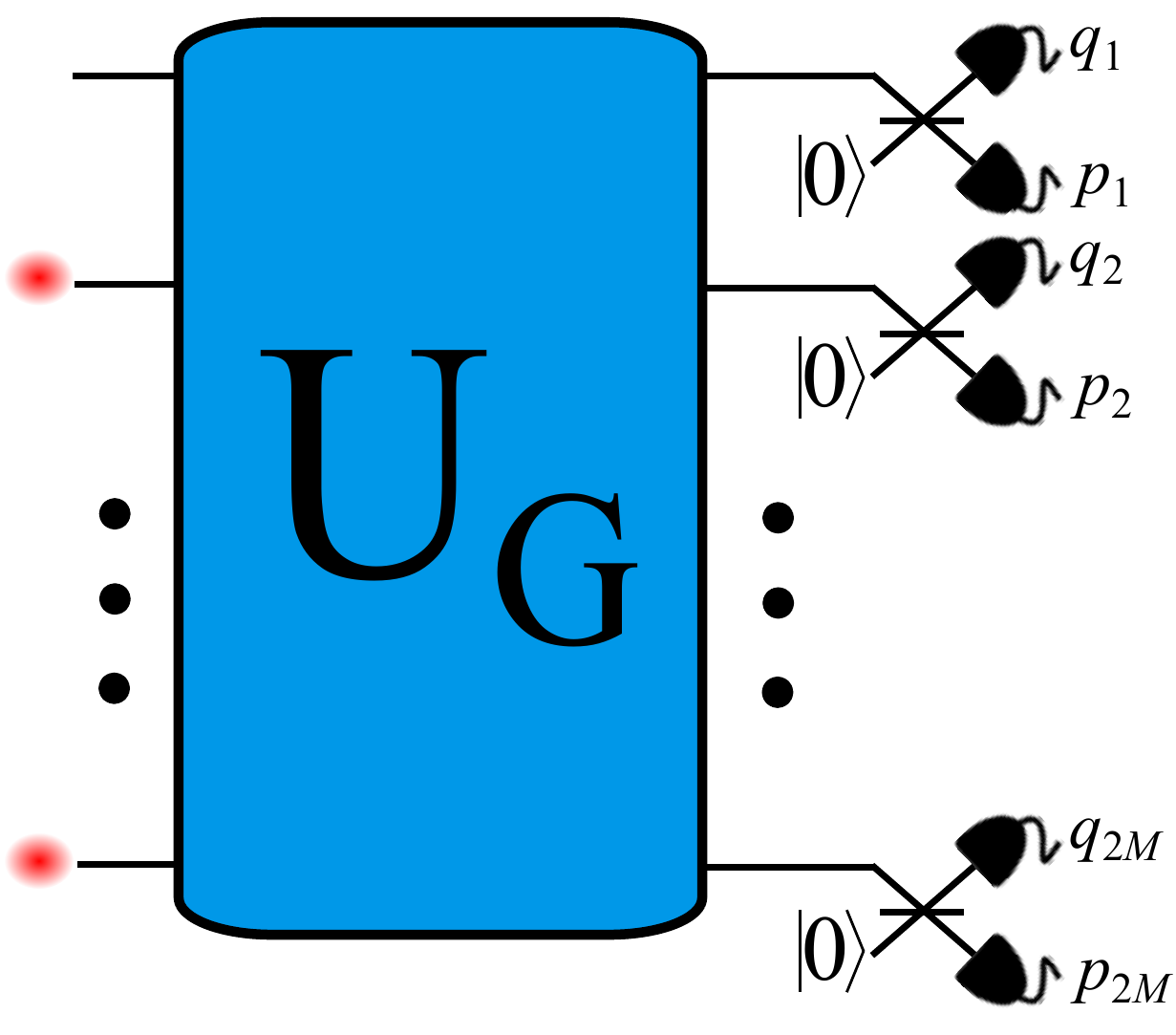}
		\caption {Boson sampling with non-Gaussian single photon input states and Gaussian eight-port homodyne detection \cite{homo}. The output at each mode of the circuit is mixed with the vacuum on a beam splitter of a reflectivity $\rho_i$, followed by the measurement of the quadratures $q_i$ and $p_i$ at each emerging port. \label{gaussian}}
	\end{center}
\end{figure}
We first note that  boson sampling with squeezed vacuum states as defined in Sec.~\ref{squeezedd}  is an instance of a classically hard task with Gaussian input states and non-Gaussian measurements. Therefore, one anticipates that a time-reversed version of this task, which would operate on non-Gaussian input states that are submitted to Gaussian measurements, should also constitute a computationally hard problem. In this section, we develop such a boson sampling model with single-photon states injected into a linear-optical circuit and followed by eight-port homodyne detection (see Fig.~\ref{gaussian}). 

More specifically, we assume that each output mode $i$ of the $2M$-port optical interferometer U$_\ttt{G}$ is mixed with the vacuum state $\ket{0}$ on a beam splitter of reflectivity $\rho_i$ and transmissivity $\tau_i$  ($\tau_i^2+\rho_i^2=1$). Next, at the emerging port of the $i$th beam splitter, the rescaled quadratures $\hat{q}_i=\hat{Q}_i/\tau_i$ and $\hat{p}_i=\hat{P}_i/\rho_i$ are measured, where $\hat{b}_i^\ttt{G}=(\hat{Q}_i+i \hat{P}_i)/\sqrt{2}$, with $\hat{b}_i^\ttt{G}$ being the $i$th output mode annihilation operator [see also Eq.~(\ref{1})]. This measurement projects the state of each output mode $i$ onto a displaced squeezed vacuum state~\cite{ulf},
\be\label{7}
\ket{\alpha_{0_i}, \xi_i}=D(\alpha_{0_i})S(\xi_i) \ket{0}.
\ee
Here, $D(\alpha_{0_i})=\exp\left(\alpha_{0_i}^{} \hat{b}_i^\ttt{G}+\alpha_{0_i}^* \hat{b}_i^{\ttt{G}^\dagger}\right)$ and $S(\xi_i)=\exp\left[\xi_i/2 \left(\hat{b}_i^{\ttt{G}^2}+\hat{b}_i^{\ttt{G}^{\dagger^2}}\right)\right]$ are, respectively, the displacement and squeezing operators, while $\alpha_{0_i}=(q_i+i p_i)/\sqrt{2}$ and $\xi_i=\ln\tau_i/\rho_i$ stand for the displacement and squeezing parameter of the measured state. The POVM elements of such a homodyne detection~\cite{homo} are projectors onto pure Gaussian states and thus it constitutes a specific example of Gaussian measurements (it is important to note that the measurement can be destructive).

Next, we denote by $\tilde{p}_{\pmb{\xi}}(\pmb{\alpha}_0|{\bf k}\cap {\bf m})$ the probability of detecting a set of squeezed displaced states $\ket{\alpha_{0_i}, \xi_i}$, given the $2M$-mode Fock input state $\ket{{\bf k}\cap {\bf m}}$ entering the interferometer U$_\ttt{G}$, where $\pmb{\xi}=\{\xi_1, \dots, \xi_{2M}\}$, $\pmb{\alpha}_0=\{\alpha_{0_1}, \dots, \alpha_{0_{2M}}\}$ and ${\bf k}\cap {\bf m}=\{k_1, \dots, k_M, m_1, \dots, m_M\}$, $k_i, m_i\in \{0,1\}$ (remark that given the linear-optical evolution of the Fock states, we expect that the function $\tilde{p}_{\pmb{\xi}}(\pmb{\alpha}_0|{\bf k}\cap {\bf m})$ is analytical in $\alpha_{0_i}$). Then, having in mind the hardness proof of the previous section, we assume that the absolute values of the squeezing parameters of states $\ket{\alpha_{0_i}, \xi_i}$ are equal, while their signs alternate, i.e., $\pmb{\xi}=\pmb{\xi}_0\equiv\{\xi, -\xi, \dots, \xi, -\xi\}$. This can be achieved by choosing the reflectivities $\rho_i$ and $\tau_i$ such that $\rho_i=\tau_{i+1}$ ($1\leq i< 2M$). We also consider unitary transformations of the type discussed in Sec.~\ref{squeezedd}, setting U$_\ttt{G}$=U$^\dagger$ [with U defined as in Eq.~(\ref{5.1.1})]. In such a case, we find 
\be \label{8}
\tilde{p}_{\pmb{\xi}_0}(0, \dots, 0|{\bf k\cap {\bf m}})=(1-t^2)^M t^{2N} |\ttt{Perm U}_{{\bf k}, {\bf m}}|^2.
\ee 
Having established this relation, we are now ready to prove that the task of sampling the displacement $\pmb{\alpha_0}$ of the measured states $\ket{\alpha_{0_i}, \xi_i}$, according to the probability distribution $\tilde{p}_{\pmb{\xi}_0}(\pmb{\alpha_0}|{\bf k}\cap {\bf m})$ is a classically hard problem. 

Following the standard procedure (see, e.g., Ref.~\cite{boxes}), we start with discretizing the phase space of every output mode into boxes of size $\eta$. More precisely, we define segments $w_j=(j \sqrt{\eta}/2,(j+2)\sqrt{\eta}/2]$ ($w_j=(-j-1) \sqrt{\eta}/2,(-j+1)\sqrt{\eta}/2]$) for odd (even) $j\geq 0$ and label each box in terms of integers $r_j$ and $s_j$, such that $q_j \in w_{r_j}$ and $p_j\in w_{s_j}$. To each box we then associate the probability 
\be\label{8.1}
\tilde{p}_{\pmb{\xi}_0}({\bf r}, {\bf s}|{\bf k}\cap {\bf m})=\prod_{i,j=1}^{2M}\int_{w_{r_i}}\int_{w_{s_j}}dp_i\, dq_j\, \tilde{p}_{\pmb{\xi}_0}(\pmb{\alpha_0}|{\bf k}\cap {\bf m}),
\ee
which corresponds to the discretization of $\tilde{p}_{\pmb{\xi}_0}(\pmb{\alpha_0}|{\bf k}\cap {\bf m})$. 

Further, following the same reasoning as in Ref.~\cite{boson}, we assume that there exists an oracle $\mathcal{O}$, which, given the description of the boson sampling circuit U$_\ttt{G}$, the squeezing parameter $\xi$ and a random string $l$ (as its {\it only} source of randomness), outputs a sample $\{{\bf r}, {\bf s}\}$ according to the distribution $\tilde{p}_{\pmb{\xi}_0}({\bf r}, {\bf s}|{\bf k}\cap {\bf m})$. The probability $p_0\equiv \tilde{p}_{\pmb{\xi}_0}({\bf r}_0, {\bf s}_0|{\bf k}\cap {\bf m})$ that $\mathcal{O}$ outputs ${\bf r}_0={\bf s}_0\equiv\{0, \dots, 0\}$ is then given as
\be\label{9}
p_0=\underset{l}{\ttt{Pr}}[\mathcal{O}(\ttt{U}_\ttt{G}, \xi, l)=\{{\bf r}_0, {\bf s}_0\}].
\ee

Next, one can relate $p_0$ to the matrix permanent of Eq.~(\ref{8}). To do so, we perform Taylor expansion of $\tilde{p}_{\pmb{\xi}_0}(\pmb{\alpha_0}|{\bf k}\cap {\bf m})$ around $\pmb{\alpha}_0=\{0, \dots, 0\}$ and plug it into the expression (\ref{8.1}), along with ${\bf r}={\bf r}_0$ and ${\bf s}={\bf s}_0$. Assuming that $\eta$ is sufficiently small, we keep terms up to the second order in the series expansion of $\tilde{p}_{\pmb{\xi}_0}(\pmb{\alpha_0}|{\bf k}\cap {\bf m})$, yielding 
\ben \label{9.1} \nonumber
p_0=&& \eta^{2M} \tilde{p}_{\pmb{\xi}_0}(0, \dots, 0|{\bf k}\cap {\bf m}) +  \\
&& \frac{\eta^{2M+2} }{24} \sum_{i,j=1}^M\frac{\partial^2}{\partial q_i \partial p_j}\tilde{p}_{\pmb{\xi}_0}(0, \dots, 0|{\bf k}\cap {\bf m})+\dots
\een
Remark that by making use of Stockmeyer's algorithm~\cite{stock}, the probability $p_0$, given the oracle $\mathcal{O}$, can be approximated to within a multiplicative error in the third level of the polynomial hierarchy.  
In turn, given the ratio $\sum_{i,j=1}^M\frac{\partial^2}{\partial q_i \partial p_j}\tilde{p}_{\pmb{\xi}_0}(0, \dots, 0|{\bf k}\cap {\bf m})/\tilde{p}_{\pmb{\xi}_0}(0, \dots, 0|{\bf k}\cap {\bf m})$, one can always choose the discretization step $\eta=2^{-\mathrm{poly}(M)}$ such that this estimate for $p_0$ translates into a polynomial-sized multiplicative error approximation for $\tilde{p}_{\pmb{\xi}_0}(0, \dots, 0|{\bf k}\cap {\bf m})$. Therefore, a classical oracle that samples from the probability distribution (\ref{8.1}) would allow one to approximate $\tilde{p}_{\pmb{\xi}_0}(0, \dots, 0|{\bf k}\cap {\bf m})$ to within a multiplicative error in the third level of the polynomial hierarchy.

Now, following Ref.~\cite{boson}, we know that one can efficiently encode a given $N\times N$ matrix X of real elements into the unitary U, so that X appears as a submatrix of U, situated, e.g., in its upper left corner. Consequently, according to Eq.~(\ref{8}), if $k_i=m_i=1$ for $1\leq i\leq N$ and $k_i=m_i=0$ for $N<i\leq M$, we find 
\be\label{10}
p_0=\varepsilon^{2 N}\ttt{Perm X}^2,
\ee
where $\varepsilon=1/||\ttt{X}||$ and $||\ttt{X}||$ is the norm of X (see Lemma 4.4 of Ref.~\cite{boson}). Therefore, combining the $\#$P hardness of estimating $\ttt{Perm X}^2$ to within a multiplicative error with the above described approximation scheme based on Stockmeyer's algorithm, we end up with a collapse of the polynomial hierarchy to its third level~\cite{boson, scattershot}. As the latter is highly unlikely, we conclude that a classical oracle $\mathcal{O}$  for sampling {\it exactly} from the distribution $\tilde{p}_{\pmb{\xi}_0}({\bf r}, {\bf s}|{\bf k}\cap {\bf m})$ does not exist. The developed model of boson sampling therefore constitutes a novel class of computationally hard tasks, involving non-Gaussian input states and Gaussian measurements.




\section{Conclusion}\label{concl}

We have proposed a model of boson sampling with Gaussian measurements which is classically hard to simulate. Our scheme is to be compared with the original and Gaussian boson sampling models, which deal, respectively, with (non-Gaussian) single-photon and (Gaussian) squeezed input states, but both involve a (non-Gaussian) discrete-outcome measurement by means of photon counting. By constructing an explicit setup involving eight-port homodyne detection, we give a positive answer to the open question whether a boson sampling task with single-photon input states and Gaussian measurements may be found that remains hard.

The model we construct and the tools we use for its hardness proof rely on the symmetry of quantum evolution under time reversal. Namely, we first modify the scattershot boson sampling model. In its original version, scattershot boson sampling operates on two-mode squeezed vacuum states, with one mode of each state being used for photon heralding, while the remaining set of modes undergo a linear-optical transformation. In contrast, in the twofold scattershot setup, both halves of two-mode squeezed states are subject to linear scattering, thus enlarging the set of classically hard photonic sampling tasks. The hardness proof of this variant of the scattershot boson sampling is established by making use of partial time reversal. The latter allows one to reveal the equivalence of the proposed scheme with the original model of Aaronson and Arkhipov.

Further, since a two-mode squeezed state can be obtained from a pair of squeezed vacuum states by mixing them on a beam splitter, our twofold scattershot scheme yields as a straightforward corollary a proof of the hardness of boson sampling from squeezed vacuum states. It generalizes the corresponding result associated with the original scattershot boson sampling model, but requires an equal degree of squeezing of all input states. Interestingly, a more general result on boson sampling from squeezed states was independently reported in Ref.~\cite{GBS}, with a more complexity-theoretic oriented proof, which nevertheless requires the same condition of equal squeezing. In fact, certain experimental setups may not allow one to produce perfectly equally squeezed-vacuum states, involving, e.g., a specific distribution of squeezings over the modes \cite{spopo_theor, spopo_exp}, while experimental imperfections might, in turn, result in a certain noise in the value of squeezing. Nevertheless, our worst-case statement that sampling squeezed states is a computationally hard problem remains valid (equal squeezing is a special case of the aforementioned scenario).

It is also worth noting that in the light of recent developments in squeezed-state generation and manipulation techniques~\cite{spopo_exp, spopo_new, furasawa}, the above  settings may find feasible implementations. Namely, our scheme allows for fewer constraints in the choice of the corresponding circuit, with a potential to fit the class of achievable unitary transformations. Specifically, a synchronously pumped optical parametric oscillator is capable of producing $\sim100$ ``supermode" squeezed states, while the equality of their squeezing parameters can be achieved by operating upon the low-order supermodes~\cite{spopo_theor}.

Finally, based on the time-reversed model of boson sampling with squeezed states, we arrive at the goal of the paper, namely the construction of a boson sampling model with non-Gaussian input states, linear optical (passive Gaussian) evolution, and Gaussian measurements. Note that any Gaussian measurement (its POVM elements being projectors onto Gaussian pure states) can be implemented with linear optics, vacuum ancillas and quadrature measurements. In our proof, we deal with a specific example, namely, eight-port homodyne detection, thus showing an explicit setting of  computationally hard boson sampling with Gaussian measurements. Further, since the Gaussian evolution of Gaussian states concluded by Gaussian measurements can be efficiently simulated on a classical computer, non-Gaussian resources are well known to be required to achieve quantum advantage~\cite{knill}. In this respect, our construction completes the set of {\it minimal} non-Gaussian extensions of Gaussian models, where, in previous works, non-Gaussianity was introduced either at the evolution or at the measurement stage (e.g., boson sampling from a Gaussian state~\cite{ralph, scattershot, GBS}, continuous-variable instantaneous quantum computing~\cite{CVI}). Note that our result proves the hardness of {\it exact} boson sampling with Gaussian measurements, but we will address the hardness of the approximate sampling case in a future work. 


In conclusion, we have illustrated how the notion of time reversal can contribute to the development of new physically-motivated computational problems and provide relatively simple tools for assessing their complexity. As a side result, the time-reversal symmetry further suggests that exact boson sampling of {\it displaced} squeezed states by means of single-photon detections is classically hard as well. We therefore believe that our approach will motivate further studies of alternative continuous-variable boson sampling models involving Gaussian resources, contributing to unveil the hierarchical structure of the computational complexity of linear optics.

{\it Note}: Recently, we became aware that related approaches to this problem have been described in Refs.~\cite{new1,new2}.


\section*{Acknowledgments }

The authors thank Giulia Ferrini and Tom Douce for useful discussions and comments. This work was supported by the H2020-FETPROACT-2014 Grant QUCHIP (Quantum Simulation on a Photonic Chip; grant agreement no. 641039).


\begin{thebibliography}{100}
\bibitem {homo} In fact, in its practical realization, this measurement scheme appears like two homodyne apparata, with four input fields involved and four output beams traveling toward four photodetectors (see also Ref.~\cite{ulf}). In view of this, the measurement is called an eight-port homodyne detection, even though some authors use the term ``heterodyne detection".
\end{thebibliography}
\end{document}